\documentclass[conference]{IEEEtran}
\IEEEoverridecommandlockouts
\usepackage[noend]{algpseudocode} 
\usepackage[]{mdframed}
\usepackage{cite}
\usepackage{amsmath,amssymb,amsfonts}
\usepackage{algorithm}
\usepackage{booktabs}
\usepackage{setspace} 
\usepackage{array}
\usepackage{multirow}
\usepackage{makecell}
\usepackage{adjustbox}
\usepackage{lipsum}
\usepackage{graphicx}
\usepackage{textcomp}
\usepackage{xcolor}
\usepackage{longtable}
\usepackage{hyperref}
\usepackage{booktabs}
\usepackage{geometry}
\usepackage{framed}
\definecolor{shadecolor}{RGB}{235,235,235}

\def\BibTeX{{\rm B\kern-.05em{\sc i\kern-.025em b}\kern-.08em
    T\kern-.1667em\lower.7ex\hbox{E}\kern-.125emX}}
\begin{document}

%Revised title as suggested by KI
\title{End-to-End Portfolio Optimization with Hybrid Quantum Annealing\\} 

%Some potential titles: 
% 1) Optimizing Portfolios with Quantum Annealing: An End-to-End Approach

% 2) Optimizing Portfolios with Quantum Annealing Techniques

% 3) Quantum Annealing in Portfolio Optimization: A Complete Solution

\author{
    Sai Nandan Morapakula\textsuperscript{1,2 *},
    Sangram Deshpande\textsuperscript{1,3 *} \thanks{ * These authors contributed equally to this work.  Contact address: S.Morapakula001@umb.edu}, 
    Rakesh Yata\textsuperscript{1},\\
    Rushikesh Ubale\textsuperscript{1},
    Uday Wad \textsuperscript{1},
    Kazuki Ikeda \textsuperscript{2,4} 
    \\[2ex]
    
    \textsuperscript{1}Qkrishi Quantum Pvt Ltd, Gurgaon, Haryana, India\\
    \textsuperscript{2}Department of Physics, University of Massachusetts, Boston, MA, USA \\
    \textsuperscript{3}Department of Electrical and Computer Engineering, North Carolina State University, Raleigh, NC, USA\\
    \textsuperscript{4}Center for Nuclear Theory, Department of Physics and Astronomy, Stony Brook University, NY, USA
}

\maketitle
% \begin{abstract}
% With rapid technological progress reshaping the financial industry, quantum technology plays a critical role in advancing risk management, asset allocation, and financial strategies. Realizing its full potential requires overcoming challenges like quantum hardware limits, algorithmic stability, and implementation barriers. This research explores integrating quantum annealing with portfolio optimization, highlighting quantum methods' ability to enhance investment strategy efficiency and speed. Using hybrid quantum-classical models, the study shows combined approaches effectively handle complex optimization better than classical methods. Empirical results demonstrate that while the individual investor’s portfolio grew from 1.6 million to 1.95 million Indian Rupees which is an increase of 350,000 Rupees or about 21.9\%—the algorithm-driven portfolio rose to 2.2 million Rupees, gaining 600,000 Rupees or 37.5\%. This represents an outperformance of approximately 250,000 Rupees, or 15.6 percentage points, over the investor’s results. Additionally, using rebalancing leads to a portfolio that also surpasses the benchmark value.

\begin{abstract}
% \textcolor{blue}
Hybrid quantum-classical optimization has emerged as a promising direction for addressing financial decision problems under current quantum hardware constraints. 
In this work we present a practical end-to-end portfolio optimization pipeline that combines (i) a continuous mean-variance and Sharpe-ratio formulation, 
(ii) a QUBO/CQM-based discrete asset selection stage solved using D-Wave’s hybrid quantum annealing solver, 
(iii) classical convex optimization for computing optimal asset weights, and 
(iv) a quarterly rebalancing mechanism.
Rather than claiming quantum advantage, our goal is to evaluate the feasibility and integration of these components within a deployable financial workflow. We empirically compare our hybrid pipeline against a fund manager in real time and indexes used in Indian stock market. The results indicate that the proposed framework can construct diversified portfolios and achieve competitive returns. We also report computational considerations and scalability observations drawn from the hybrid solver behaviour. While the experiments are limited to moderate sized portfolios dictated by current annealing hardware and QUBO embedding constraints, the study illustrates how quantum assisted selection and classical allocation can be combined coherently in a real-world setting. This work emphasizes methodological reproducibility and practical applicability, and aims to serve as a step toward larger-scale financial optimization workflows as quantum annealers continue to mature.
\end{abstract}

% \end{abstract}


\section{Introduction}
%\ki{Overall, minimize all kinds of subjective works, like greatly, quite, significant, etc... Use these only when to emphasize something extremely important. Keep the sentences scientific, avoid hypo, and focus on facts.}
In the evolving landscape of financial technology, the task of optimizing investment portfolios presents both challenges and opportunities. Traditional methods for portfolio optimization, while well-established, often struggle under the complexity and large data requirements of modern financial markets. As the demand for faster and more efficient solutions grows, the integration of quantum technology into financial strategies emerges as a new approach. This paper examines how classical and quantum methods can be combined to enhance the process of financial portfolio optimization, focusing on the intricacies of modern investment environments.

Portfolio optimization is a core financial task that seeks to allocate assets in a manner that maximizes returns while minimizing risk. Traditionally, this process relies on mathematical models such as Modern Portfolio Theory (MPT)\cite{mpt} and Mean-Variance Optimization (MVO)\cite{omf}, which use classical algorithms to assess and predict the behavior of various asset classes. However, these classical solutions often encounter limitations in scalability and speed\cite{https://doi.org/10.1111/1540-6261.00580}, \cite{ffa6df70-e927-37dd-b244-18163d29888c}, \cite{LEDOIT2004365}, \cite{chen2013sparseportfolioselectionquasinorm}, particularly when dealing with large datasets or complex asset interactions.

Quantum methods, characterized by the potential to process information at scales beyond conventional technology limits, offers an alternative framework to address these challenges. By leveraging principles of quantum mechanics, such as superposition and entanglement, quantum algorithms can search extensive solution spaces more efficiently, potentially identifying improved portfolio allocations with greater speed and accuracy. Furthermore, the emergence of hybrid models that combine classical and quantum techniques provides a balanced approach, optimizing asset selection with quantum algorithms while using classical methods for asset allocation and risk assessment.

This paper outlines these methodologies and provides both empirical analysis and theoretical insights into their effectiveness. By integrating quantum annealing into portfolio optimization \cite{Lucas_2014,glover2019tutorialformulatingusingqubo}, we aim to establish a more adaptive and efficient framework that can operate in the dynamic and often unpredictable realm of financial markets.

The specific achievements we produced through our results are summarized as follows:
\begin{shaded}
\begin{itemize}
    \item The hybrid quantum classical model was able to allocate better weights to the stocks compared to a fund manager in real life, as presented in  Fig.~\ref{fig:portfolio_comparison}.
    \item Rebalancing using the fully quantum version improved the bank's portfolio, as observed in  Fig.~\ref{fig:rebalancing_comparison}.
\end{itemize}
\end{shaded}

% \textcolor{blue}{
This study does not claim that quantum annealing or the hybrid CQM solver provides superior runtime or scalability relative to tuned classical solvers. 
We refrain from drawing conclusions about asymptotic scaling or computational speed. 
The contribution of this work lies in demonstrating a feasible, end-to-end integration of quantum-assisted discrete selection with classical allocation and rebalancing on real financial data, rather than establishing quantum computational advantage.
% }

In the remainder of this introduction, we review related work in finance using quantum hardware (Sec.~\ref{sec:background}). Next, we present the methodology (Sec.~\ref{sec:meth}) employed to obtain our findings. In Sec.~\ref{sec:results}, we discuss the results in detail, and finally, we conclude the paper in Sec.~\ref{sec:conc}.

% \begin{figure*}
%     \centering
%     \includegraphics[width= \textwidth]{flowchart_diagram1.png}
%     \caption{Flowchart of the Algorithm}
%     \label{fig:f1}
% \end{figure*}

\section{Background and Related Work}\label{sec:background}

Quantum annealing (QA)~\cite{PhysRevE.58.5355} has emerged as a practical approach for solving combinatorial optimization problems in finance.
D-Wave Systems has developed several generations of quantum annealers, including the 2000Q\textsuperscript{TM} and the Advantage family, which employ transverse-field Ising architectures for energy minimization~\cite{Hauke_2020, Abbas_2024, Yarkoni_2022, Dwave-whitepaper}.
These systems implement optimization problems formulated as Quadratic Unconstrained Binary Optimization (QUBO) or Ising models, following the early mappings proposed by Rosenberg~\cite{Rosenberg_2016} and Glover~\cite{glover2019tutorialformulatingusingqubo}.
Recent work has also extended this framework to polynomial unconstrained binary optimization (PUBO)~\cite{grange2024quadraticversuspolynomialunconstrained}.

Applications of QA to financial modeling began with Venturelli and Kondratyev~\cite{Venturelli_2019}, who addressed multi-period portfolio optimization with transaction costs on physical D-Wave hardware, demonstrating early scalability.
Subsequent studies have adopted both annealing and gate-based paradigms.
The Quantum Approximate Optimization Algorithm (QAOA)~\cite{farhi2014quantumapproximateoptimizationalgorithm} has been explored for portfolio allocation and rebalancing~\cite{hodson2019portfoliorebalancingexperimentsusing, huot2024enhancingknapsackbasedfinancialportfolio}, though results remain hardware-limited.
Comparative benchmarks~\cite{Brandhofer_2022, mazumder2024benchmarkingmetaheuristicintegratedqaoaquantum, zaman2024poqaframeworkportfoliooptimization} indicate that quantum methods currently perform comparably to classical heuristics such as simulated annealing and genetic algorithms, with hardware precision and connectivity constraints being the main bottlenecks.
Hybrid strategies that combine classical optimizers with quantum subroutines are therefore viewed as the most feasible near-term route~\cite{Egger_2020}.
Related advances include decomposition-based hybrid frameworks for large portfolios~\cite{acharya2024decompositionpipelinelargescaleportfolio} and hybrid eigensolver models~\cite{wang2024variationalquantumeigensolverlinear}.

Beyond portfolio optimization, quantum techniques have been applied to risk analysis, option pricing, and arbitrage detection.
For instance, Woerner and Egger~\cite{Woerner_2019} employed amplitude estimation to accelerate Value-at-Risk (VaR) and conditional VaR computations, while Stamatopoulos et al.~\cite{Stamatopoulos_2020} demonstrated quantum-based option pricing speedups.
Orús et al.~\cite{Or_s_2019} provided a comprehensive survey of such applications, emphasizing both the promise and the practical limitations of current devices.

% \textcolor{blue}{
While several prior works have explored hybrid quantum–classical approaches for portfolio optimization~\cite{Venturelli_2019, Rebentrost_2018, Mugel_2022},
most studies remain constrained by simplified datasets, limited covariance structures, and a lack of dynamic rebalancing.
Few integrate the entire pipeline from mean–variance formulation through QUBO mapping, annealing, decoding, and time-series-based rebalancing within a unified financial framework.
Our work addresses these gaps by presenting a comprehensive end-to-end quantum annealing model that
(1) reformulates continuous mean–variance and Sharpe-ratio objectives into QUBO form without ad-hoc linearization,
(2) employs D-Wave’s hybrid solver to capture real-market correlations from the Indian equity market, and
(3) demonstrates reproducible rebalancing that outperforms the fund manager in both return and risk metrics.
% }

\section{Methodology}\label{sec:meth}
In this section, we elucidate the problem formulation methodologies employed in this study. A prominent technique is quantum annealing, which leverages the adiabatic theorem. According to this theorem, a quantum system remains in its ground state provided that the Hamiltonian evolves sufficiently slowly over time. Although practical implementations may not fully satisfy the adiabatic condition, they still approximate it closely enough to be useful in practice.
To address an optimization problem, a time-dependent Hamiltonian \( H(t) \) is constructed. Initially, the system is prepared in the ground state of an easily realizable Hamiltonian \( H_0 \). Over time, the system evolves towards the target Hamiltonian \( H_1\), with its ground state encapsulating the solution. The gradual transition is defined as follows \cite{PhysRevE.58.5355}:
\begin{equation}
H(t) = \left(1 - \frac{t}{\tau}\right) H_0 + \frac{t}{\tau} H_1,~0\le t\le\tau,
\end{equation}
where \(\tau \) represents the entire annealing time.  According to the adiabatic theorem, if this evolution is slow enough, the system will remain in its instantaneous ground state throughout, resulting in the ground state of \(H_1 \) when \(t = \tau \).  Typically, the initial Hamiltonian is defined as \(H_0 = -\sum_i X_i \), where \(X_i \) represents the Pauli-X operator acting on the \(i \)-th qubit.  The problem Hamiltonian takes the form of an Ising model:

\begin{equation}
H_1 = \sum_{i > j} J_{ij} Z_i Z_j + \sum_i h_i Z_i,
\end{equation}
where \( J_{ij} \) represents couplings between qubits \( i \) and \( j \), \( h_i \) denotes external fields, and \( Z_i \) is the Pauli-Z operator.

This principle is realized by commercial quantum annealers, such as those created by D-Wave Systems, which use quantum processing units (QPU).

Specifically, we will utilize the constrained quadratic model (CQM). The CQM solver can handle problems with binary and integer variables and supports both equality and inequality constraints. It is reported that the CQM solver demonstrates superior performance in solving problems characterized by a large number of constraints, compared to other hybrid approaches like the Binary Quadratic Model (BQM) and the Discrete Quadratic Model (DQM) \cite{cqm-dwave-whitepaper}. In this work, we have used the CQM solver (hereinafter referred to as the hybrid solver), hybrid\_constrained\_quadratic\_model\_version1, provided by D-Wave to simulate the problem. This approach is better suited for constraint-based issues than the conventional QUBO formulation used in quantum annealing.  

\subsection{Modern Portfolio Theory (MPT) formulation}
Based on \cite{mpt}, for $n$ stocks with average monthly return $\mu_i$ per dollar spent on stock $i$ and covariance $\sigma_{ij}$ between stocks $i$ and $j$, the portfolio optimization problem, given a budget of $B$ dollars, involves determining the optimal number of shares $x_i$ of each stock $i$ purchased at price $p_i$ per share at risk $q>0$. This can be formulated as: 
\begin{align}
     \label{eq:Non-negativity constraint}
     \min &~ q\left( \sum_{i=1}^n \sum_{j=1}^n \sigma_{ij}(p_ix_i)(p_jx_j)\right) - \left( \sum_{i=1}^n\mu_i(p_ix_i) \right) \\
    & \text{subject to:}~ \sum_{i=1}^n(p_ix_i) \leq B\label{eq:budgetconstraint} 
\end{align}
%\si{take a look again at MPT formulation}

Eq.~\eqref{eq:budgetconstraint} is the budget constraint which means that a user cannot spend more than the budget and it should always be less than or equivalent to it. Likewise, eq.~\eqref{eq:Non-negativity constraint} is a non-negative constraint meaning that the user cannot buy negative shares of any stock simply put, number of shares of any stock should always be positive.
Here, $q > 0$ is the risk aversion coefficient i.e. the trade-off coefficient between the risk (variance) and the returns.

The number of shares \( x_i \) affects the portfolio risk through its contribution to the overall variance term. Since the variance is calculated using the covariances \( \sigma_{ij} \), the individual standard deviation \( \sigma_i = \sqrt{\sigma_{ii}} \) indirectly influences the portfolio's total risk depending on the amount invested in asset \( i \), i.e., \( p_i x_i \). Thus, higher allocations to more volatile stocks (with high \( \sigma_i \)) increases the portfolio variance.

%\ki{Describe the relation between $x_i$ and $\sigma^z_i$.}

\subsection{Sharpe ratio maximization}

A classical method usually maximizes the Sharpe ratio, a measure for the performance of investments \cite{omf}:
\begin{equation}
    \text{Sharpe ratio} = \frac{E[R_p - R_f]}{\sigma_p} = \frac{E[R_p - R_f]}{\sqrt{var[R_p - R_f]}}, \label{eq:sharperatio}
\end{equation}
where $R_p$ is the return of the portfolio, $R_f$ is the risk-free rate, $E[R_p - R_f]$ is the expected value of excess of portfolio returns and $\sigma_p$ is standard deviation of the portfolio's excess return.

For simplicity, we take $R_f = 0$ and substitute it in eq.~\eqref{eq:sharperatio}. We get, 

\begin{align}
    & \text{Sharpe ratio} = \frac{E[R_p]}{\sqrt{var[R_p]}} = \frac{\mu^T x}{\sqrt{x^T \Sigma x}}, ~~~~ x \in \mathbb{R}^n.
\end{align}

A Sharpe ratio above 1.0 is deemed good, with 2.0 being very good and 3.0 or higher considered excellent. Ratios below 1.0 are considered sub-optimal.

The above formulation is non-convex.  In general, non-convex problems are very difficult to solve. Fortunately, there exists a convex reformulation i.e.

\begin{equation}
\begin{aligned}
\min \quad & y^T\,\Sigma\,y,\\
\text{s.t.} \quad & (\mu - r)^T\,y \;=\; 1 \text{and }  y \;\ge\; 0.
\end{aligned}
\label{eq:convex_ref}
\end{equation}

We consider $r_f = 0$. Then map the optimal solution $y^*$ back to the original problem  via the transformation.

\begin{align*}
    w^*_i &:= \frac{y^*_i}{\sum_{j=1}^n y^*_j} \qquad i = 1, \ldots, n.
\end{align*}

%Current idea: Use the convex formulation to find the best portfolio $y^*$ which gives the sum $\sum_{j=1}^n y^*_j = k$. Use this value to impose cardinality constraint in a new experiment.

We first use the classical convex formulation, which does not impose any explicit cardinality requirements, to practically implement a cardinality constraint in our portfolio optimization. The optimal continuous portfolio vector \( y^* \) is obtained by solving this convex optimization. Through the explicit computation of \[
    k = \sum_{j=1}^{n} y_j^*.
\] We derive the desired cardinality from this solution by assessing the sum of its constituents. 
 The optimal number of assets is indicated by the resulting value \( k \), which offers a well-informed and theoretically supported cardinality restriction.  Then, in a new discrete or quantum-based optimization experiment, we explicitly impose this derived cardinality \( k \). 

\begin{align*}
    0 \leq y \leq kx \\
    1^Tx = \kappa
\end{align*}

We combine the computational efficiency and robustness of convex optimization with the usefulness of discrete cardinality constraints appropriate for quantum or combinatorial optimization techniques by using this two-step approach, which involves first solving a convex formulation to determine an optimal cardinality and then using that cardinality as a fixed parameter in a subsequent optimization.  This hybrid approach guarantees that the finished portfolio has a theoretically sound base while adhering to real-world investment constraints.

\subsection{Mean-variance portfolio optimization (MVO)}
The MVO formulation aims to maximize the returns as well as minimize the risk \cite{omf}. For solving the problem on quantum computers, we need to formulate it as a QUBO problem.

$$\begin{aligned}
\min_{x \in \{0, 1\}^n}  q x^T C x - \mu^T x\\
\text{subject to: } 1^T x = B,
\end{aligned}$$
where we use the following notation: 

\begin{itemize}
    \item $x \in \{0, 1\}^n$ denotes the vector of binary decision variables, which indicate which assets to pick ($x[i] = 1$) and which not to pick ($x[i] = 0$)
    \item $\mu \in \mathbb{R}^n$ defines the expected returns for the assets
    \item $C \in \mathbb{R}^{n \times n}$ specifies the covariances between the assets
    \item $q > 0$ controls the risk appetite of the decision-maker
    \item and $B$ denotes the budget, i.e. the number of assets to be selected out of $n$
\end{itemize}
cov for covarience but not sigma as it is confusing 
% We need to remove the constraint and formulate the problem as a QUBO to make it suitable for quantum computer. \textcolor{blue}{
To convert the budget equality constraint \(1^\top x = B\) into an unconstrained QUBO,
we add a positive quadratic penalty term \(\lambda_b (1^\top x - B)^2\) to the objective.
For a minimization problem, violations of the constraint must increase the objective,
hence the penalty enters with a positive sign.
% }

% The equality constraint $1^T x = B$ is mapped to a penalty term $(1^T x - B)^2$ which is scaled by a parameter (Lagrange multiplier) and subtracted from the objective function. 

% \begin{equation} \label{quboformulation}
% \min_{x \in \{0, 1\}^n}  q x^T C x - \mu^T x - \lambda (1^T x - B)^2,
% \end{equation}
% \textcolor{blue}{
\begin{equation}
\label{quboformulation}
\min_{x\in\{0,1\}^n}\;
q\,x^\top C x
\;-\;
\mu^\top x
\;+\;
\lambda_b\, (1^\top x - B)^2 ,
\end{equation}
% }

where $\lambda$ is called the Lagrange multiplier which acts as a penalty term (tunable). This will serve as the hamiltonian for our simulation. The below explains the hamiltonian below elaborately: 
%\ki{describe the Hamiltonian}
\begin{itemize}
    \item \textbf{Risk term:} \( q x^T C  x \) \\
    This term represents the \emph{portfolio risk}, measured as a quadratic form over the covariance matrix. It penalizes portfolios with high variance. The scalar \( q > 0 \) reflects the investor’s \emph{risk aversion}—larger values of \( q \) prioritize risk minimization.

    \item \textbf{Return term:} \( -\mu^T x \) \\
    This term rewards portfolios with higher \emph{expected return}. The negative sign ensures that higher returns reduce the overall energy, thereby making them more favorable in the optimization.

    \item \textbf{Budget constraint penalty:} \( -\lambda_b (1^T x - B)^2 \) \\
    This is a \emph{quadratic penalty} enforcing the selection of exactly \( B \) assets. The Lagrange multiplier \( \lambda \) controls the strength of this constraint. Higher values penalize deviations from the budget more severely.
\end{itemize}

% \textcolor{blue}{
The penalty weight \(\lambda_b\) is chosen large enough that any violation of
\(1^\top x = B\) yields a higher objective than the best feasible improvement. 
Following standard practice, we initialize \(\lambda_b\) using the magnitude of the QUBO coefficients
and increase it iteratively until the constraint is satisfied within a small tolerance.
% }
We convert the QUBO problem \eqref{quboformulation} into an Ising Hamiltonian \cite{Lucas_2014}. As the Ising model has spin variables $s_i\in \{-1,1\}$, ${x_i}$ is transformed to (1+${s_i}$)/2. The classical Ising Hamiltonian of our QUBO problem is given as follows:
\begin{equation}\label{qubotoising1}
    \min_{s} \left( 
\sum_i h_i s_i + \sum_{ij} J_{ij} s_i s_j + 
\lambda \left( \sum_i \pi_i s_i - \beta \right)^2 
\right)
\end{equation}
\[
\quad \text{s.t.} \quad s_i \in \{-1, 1\} \quad \forall i,
\]
where ${J_{ij}}$ is the coupling term between two variables. To implement this Hamiltonian on a D-Wave quantum annealer, we first transform the problem into the language of the Pauli $Z$ operators, which have eigenvalues of $\pm 1$.  The Ising Hamiltonian is:
\begin{equation}\label{qubotoquantumising}
H = \sum_i h_i Z_i + \sum_{ij} J_{ij} Z_i Z_j + \lambda \left( \sum_i \pi_i Z_i - \beta \right)^2.
\end{equation}

\subsection{Hybrid Quantum-Classical Formulation}

\begin{figure}[h!]
    \centering \includegraphics[width=0.4 \textwidth]{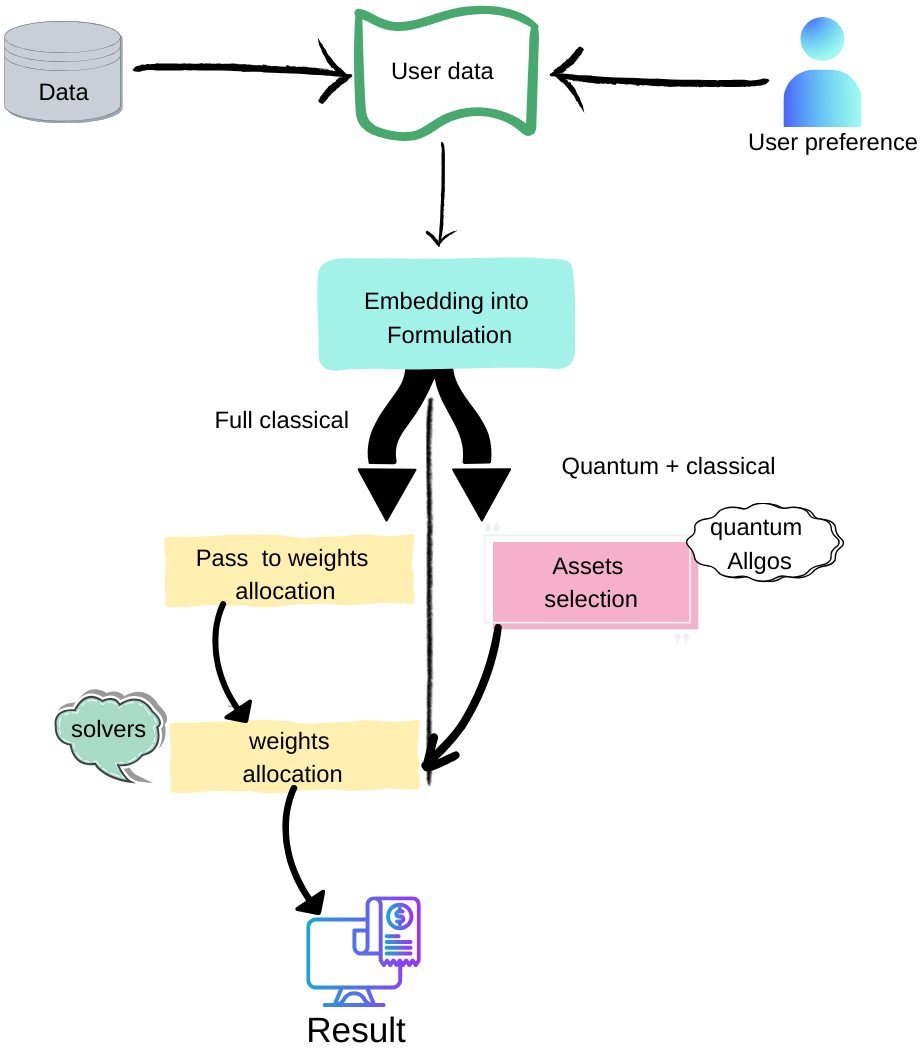}
    \caption{Hybrid Quantum Classical Flow Diagram}
    \label{fig:HQCFD}
\end{figure}

Based on the formulations above, the hybrid quantum-classical approach involves using a quantum computer for asset selection and a classical solver for asset weight allocation, as illustrated in Fig.~\ref{fig:HQCFD}. The MVO formulation is employed for selecting assets, while both MVO and the maximum Sharpe ratio (maximizing eq.~\eqref{eq:sharperatio} with respect to $R_p-R_f$) can be used for allocating weights to the selected assets.

\subsection{Rebalancing}\label{subsec:rebalancing}

The idea is to rebalance portfolio at each time period until the target time is reached. 

\subsubsection{Risk identification}
Risk identification in a portfolio involves closely analyzing the historical performance and volatility of individual assets to predict potential risks and returns. Risky companies/assets are identified by observing the expected daily returns of stocks over a period of time. This analysis not only helps in identifying which companies or assets may be riskier but also aids in understanding how these risks could impact the overall portfolio. 

\subsubsection{Portfolio Health check}

Every three months, we conduct a comprehensive review of each asset's mean returns. We monitor the portfolio value and calculate the profit. Underperforming assets, identified as those with unsatisfactory performance and elevated risk levels, are systematically removed from the portfolio. This process ensures that the portfolio remains aligned with our rebalancing specifications and maintains optimal health in regard to risk and returns.

% \subsubsection{Buy Idea}

% For identified risky companies (say N number of risky companies), all their stocks are sold as of now. The sales revenue generated is used as the new budget to purchase stocks again. 

% \begin{itemize}
%     \item Identify the sectors of the companies whose stocks are sold
%     \item Obtain the list of companies in those sectors excluding the ones whose stocks are sold.
%     \item Perform portfolio optimization using the list of companies and the new budget. The output of the algorithm is a list of $N$ companies along with the number of stocks to purchase of each company.
%     \item Buy those stocks and update the holdings.
% \end{itemize}

% \textcolor{blue}{
A detailed description of rebalancing technique is given below:\\
\textbf{Input:} initial holdings $H^{(0)}$; sector map $s(i)$; budget $B$; risk-aversion $q$; sell count $K_{\text{sell}}$.\\
\textbf{Output:} new portfolio, its portfolio value, and risk/return metrics. 
\textit{(1) Estimate returns/covariance per quarter:}
compute HP-filtered prices, log-returns on, then annualized mean $\mu^{(k)}$ and covariance $\Sigma^{(k)}$.
\textit{(2) Identify underperformers:}
rank the \emph{current} holdings in $H^{(0)}$ by $\mu^{(k)}$; mark the lowest as \emph{underperformers} and liquidate them.
\textit{(3) Sector-matched candidate set:}
collect replacements from the same sectors as the sold names, available in the universe at quarter end.
\textit{(4) Discrete selection (QUBO):}
solve equation \ref{quboformulation}
using a classical SA sampler for the QUBO to pick new assets.
\textit{(5) Integer allocation (CQM, hybrid):}
for chosen names, solve a single-period CQM on D-Wave’s hybrid solver, update holdings and wallet.
\textit{(6) Record metrics:} compute portfolio value, return, risk, and Sharpe ratio of the new portoflio.
In our implementation, “risky companies” refers \emph{not} to high-variance names but to \emph{underperformers by expected return $\mu$}.
% }
% We retain variance via $\Sigma$ in the objective and provide an ablation in the appendix varying the sell rule (lowest $\mu$ vs. highest marginal variance contribution).

\begin{figure}[h!]
    \centering
    \includegraphics[width=0.4 \textwidth]{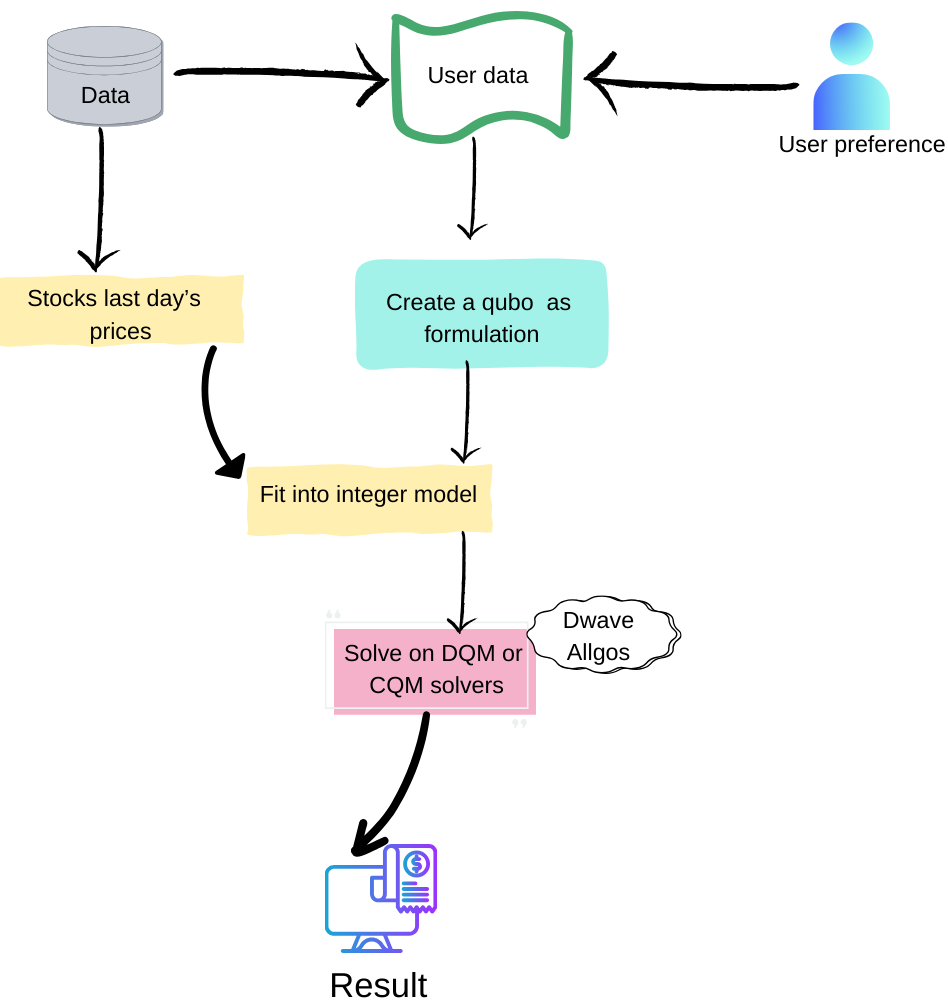}
    \caption{Fully Quantum Flow Diagram}
    \label{fig:FQFD}
\end{figure}

Fig.~\ref{fig:FQFD} illustrates the flow diagram of rebalancing, where fully quantum approach has been used in generating the results. 

% \textcolor{blue}{
Although the underlying optimization models follow the standard Modern Portfolio Theory and Mean–Variance formulations,
our implementation introduces several key innovations in translating these into a QUBO/Ising framework.
First, the continuous portfolio weights are discretized through a structured mapping that preserves convexity and avoids arbitrary thresholding.
Second, an adaptive penalty factor~$\lambda$ is applied within the constraint Hamiltonian~$H_c$
to dynamically balance the trade-off between the return–risk objective and the budget or cardinality constraints during the annealing process.
Third, after sampling from the annealer, a normalization-based decoding step ensures that the resulting bitstrings correspond to feasible weight allocations whose total sum equals the capital budget.
Finally, the Hamiltonian includes a temporal continuity term that links portfolios across rebalancing intervals, encouraging smoother transitions and reduced turnover.
These additions collectively form the methodological contribution of our work and distinguish the proposed framework from existing static or single-period QUBO formulations used in earlier quantum finance studies.
% }

\section{Results and Discussion}\label{sec:results}
In this section, we delve into the empirical results achieved through our research. We first to examine the dataset utilized in our experiments facilitated by D-Wave's Quantum Annealer. The data, reflecting real-time financial conditions, was supplied by a banking institution. Unless explicitly stated, all monetary values referenced herein are denominated in Indian Rupees (INR).

Table \ref{attributionreport} presents an overview of a representative user's portfolio at a specific point in time, which serves as a comparative framework for our analysis. For benchmarking purposes, we employ the HDFCNIFTY50 ETF, and the user's portfolio itself becomes the benchmark against which our results using the quantum annealer are evaluated. Through this meticulous comparison, we aim to validate the performance enhancements derived from our quantum approach.

\begin{table*}[t]
\centering
\caption{Attribution Report of a real world portfolio. Note: Port = Portfolio, Bench = Benchmark, Ret = Return. Start date = 1 Januray 2023, end date = 29 February 2024 Values represent percentage weights and returns.}
\label{attributionreport}
\resizebox{\textwidth}{!}{%
\begin{tabular}{@{}lrrr|rrr@{}}
\toprule
\multirow{2}{*}{\textbf{Security}} & \multicolumn{3}{c|}{\textbf{Average Weight (\%)}} & \multicolumn{3}{c}{\textbf{Tot Return (\%)}} \\
\cmidrule(lr){2-4} \cmidrule(lr){5-7}
 & \textbf{Port} & \textbf{Bench} & \textbf{+/-} & \textbf{Port Ret} & \textbf{Bench Ret} & \textbf{+/-} \\
\midrule
\textbf{NIFTY TOP 10} & 100.00 & 100.00 & 0.00 & 22.79 & 26.91 & -4.13 \\
\midrule
\textbf{Information Technology} & 28.66 & 13.64 & 15.02 & 16.61 & 19.11 & -2.49 \\
\quad TATA CONSULTANCY SVCS LTD & 20.25 & 4.17 & 16.08 & 19.70 & 19.70 & 0.00 \\
\quad INFOSYS LTD & 8.41 & 6.07 & 2.34 & 9.90 & 9.90 & 0.00 \\
\midrule
\textbf{Financials} & 18.29 & 36.37 & -18.08 & 9.05 & 8.92 & 0.13 \\
\quad HDFC BANK LIMITED & 9.36 & 11.20 & -1.85 & -11.60 & -11.60 & 0.00 \\
\quad ICICI BANK LTD & 5.52 & 7.76 & -2.24 & 25.35 & 25.35 & 0.00 \\
\quad STATE BANK OF INDIA & 3.42 & 2.64 & 0.78 & 46.71 & 46.71 & 0.00 \\
\midrule
\textbf{Consumer Staples} & 17.44 & 9.38 & 8.06 & -2.18 & 15.37 & -17.56 \\
\quad HINDUSTAN UNILEVER LTD & 14.91 & 2.68 & 12.23 & -5.16 & -5.16 & 0.00 \\
\quad ITC LTD & 2.53 & 4.49 & -1.96 & 19.20 & 19.20 & 0.00 \\
\midrule
\textbf{Industrials} & 15.88 & 5.40 & 10.48 & 63.59 & 73.15 & -9.56 \\
\quad LARSEN \& TOUBRO LTD & 15.88 & 3.84 & 12.04 & 63.59 & 63.59 & 0.00 \\
\midrule
\textbf{Energy} & 14.49 & 11.89 & 2.60 & 40.04 & 50.74 & -10.70 \\
\quad RELIANCE INDUSTRIES LTD & 14.49 & 9.89 & 4.60 & 40.04 & 40.04 & 0.00 \\
\midrule
\textbf{Communication Services} & 5.23 & 2.63 & 2.60 & 44.99 & 44.99 & 0.00 \\
\quad BHARTI AIRTEL LTD & 5.23 & 2.63 & 2.60 & 44.99 & 44.99 & 0.00 \\
\bottomrule
\end{tabular}%
}

\end{table*}

Before computing the value of the portfolio, here we assign the weights to all the given stocks. The weights are allocated based on the classical algorithm that we discussed above. Taking the weights into consideration we calculate the sharpe ratio, returns, risk and diversification ratio of the given stocks.

% \setlength{\arrayrulewidth}{0.5mm}
% \setlength{\tabcolsep}{10pt}
% \renewcommand{\arraystretch}{1.5}
% \begin{table}[!h]
% \centering
% \begin{tabular}{ |p{2cm}|p{2cm}|p{2cm}|  }
% \hline
% Name of the stock & Weights allocated by algorithm &Weights given in the benchmark \\
% \hline
% Bharti Airtel & 16.26 & 5.23 \\
% HDFC Bank & 2.59   & 9.36 \\
% Hindustan Unilever &4.59 & 14.91 \\
% ICICI Bank &10.42 & 5.52 \\
% Infosys & 6.90 & 8.41 \\
% ITC & 10.56 & 2.53   \\
% L\&T & 17.73 & 15.88 \\
% Reliance Industries & 12.42 & 14.49 \\
% State Bank of India & 10.24 & 3.42 \\
% TCS & 8.28 & 20.25 \\
% \textbf{Returns} & \textbf{26.79} & \textbf{18.03} \\
% Risk & 10.49 & 10.92 \\
% \textbf{Sharpe Ratio} & \textbf{2.55} & \textbf{1.65} \\
% Diversification Ratio & 1.78 & 1.72\\
% \hline
% \end{tabular} 
% \caption{Algo wts vs Benchmark wts}
% \label{algowts vs benchwts}
% \end{table}

\setlength{\arrayrulewidth}{0.5mm}  % Thicker rule
\setlength{\tabcolsep}{10pt}       % Column separation
\renewcommand{\arraystretch}{1.5}  % Row spacing

\begin{table}[htbp]  % Let LaTeX place it top/bottom
\centering

% In IEEE, caption typically goes ABOVE the table
\caption{Algorithm weights vs Benchmark weights}
\label{tab:algowts_vs_benchwts} % No spaces in label

\begin{tabular}{|p{2cm}|p{2cm}|p{2cm}|}
\hline
\textbf{Name of the stock} & \textbf{Weights by algorithm} & \textbf{Weights in benchmark} \\
\hline
Bharti Airtel & 16.26 & 5.23 \\
HDFC Bank & 2.59 & 9.36 \\
Hindustan Unilever & 4.59 & 14.91 \\
ICICI Bank & 10.42 & 5.52 \\
Infosys & 6.90 & 8.41 \\
ITC & 10.56 & 2.53 \\
L\&T & 17.73 & 15.88 \\
Reliance Industries & 12.42 & 14.49 \\
State Bank of India & 10.24 & 3.42 \\
TCS & 8.28 & 20.25 \\
\textbf{Returns} & \textbf{26.79} & \textbf{18.03} \\
Risk & 10.49 & 10.92 \\
\textbf{Sharpe Ratio} & \textbf{2.55} & \textbf{1.65} \\
Diversification Ratio & 1.78 & 1.72 \\
\hline
\end{tabular}
\end{table}

Table \ref{tab:algowts_vs_benchwts} presents a comparative analysis of the asset allocation weights determined by our algorithm against those provided by the bank's benchmark. This table also includes key financial metrics such as risk, return, Sharpe ratio, and diversification ratios. Our algorithm's allocation strategy results in superior financial performance. Specifically, the algorithm's weight distribution yields higher returns and a more favorable Sharpe ratio in comparison to the benchmark distribution. This demonstrates the efficacy of our method in optimizing asset allocation to achieve enhanced performance metrics.
% \textcolor{blue}{
It is important to note that the weight allocations in Table \ref{tab:algowts_vs_benchwts} are
derived from the classical MIQP optimizer,
while the quantum solver is utilized only during the rebalancing
stage. The overweighting of ICICI Bank, ITC, and Bharti Airtel
reflects their superior expected return-to-risk characteristics
within the classical covariance structure. 
% }

%\begin{figure}[h!]
    %\centering
    %\includegraphics[width=0.5\textwidth]{images/weights_algo.png}
    %\caption{Weights, Risk and Returns given by the algorithm}
    %\label{fig:weights_algo}
%\end{figure}

Once the weights are assigned, we determine the total investment amount used to purchase stocks across all the selected companies. Based on this total investment, we calculate the optimal number of shares to acquire for each company. By multiplying the optimal number of shares by the closing price of each company's stock, we aggregate these values to determine the overall portfolio value. 

Furthermore, the portfolio configured using our algorithm's assigned weights demonstrates significantly higher returns and a superior Sharpe ratio compared to the benchmark. Notably, this portfolio also exhibits reduced risk relative to the benchmark, highlighting the effectiveness of our method in optimizing investment outcomes.

 Algo \ref{algo1} summarizes the complete end-to-end workflow implemented in this study. The pipeline begins with data ingestion and preprocessing, where daily prices are converted into log-returns and used to estimate the expected-return vector and covariance matrix for each rebalancing window. These estimates feed into a discrete asset-selection module, formulated as a QUBO and solved using a quantum-assisted (or simulated) annealing routine. The selected assets are then passed to an integer-constrained CQM allocator that computes feasible share allocations under budget, sector, and cardinality constraints. Finally, a quarterly rebalancing mechanism identifies underperforming holdings, replaces them through a repeat of the selection stage, and updates the portfolio’s evaluation metrics. This unified structure provides a transparent hybrid framework that integrates quantum-assisted selection with classical optimization and realistic trading constraints.

\begin{algorithm}\label{algo1}
\caption{End-to-End Portfolio Optimization and Rebalancing via Quantum Annealing}
\begin{algorithmic}[1]
  \State \textbf{Input:} Historical prices $P_{i,t}$ for assets $\mathcal{U}$; 
         rebalancing interval $\Delta t=quaterly$; constants $q, \lambda, K_{\text{sell}}, B$
  \State \textbf{Output:} Optimized quarterly portfolios $\{w^{(k)}\}$ with risk–return metrics
  \Statex

  \State \textit{Data Preparation}
  \State Retrieve adjusted close prices for $\mathcal{U}$ from NSE/BSE between $t_{\text{start}}$ and $t_{\text{end}}$
  \State Compute log-returns, annualized mean $\mu$, and covariance matrix $\Sigma$
  \State Apply HP filter ($\lambda_{\text{HP}} = 6.25$) to smooth price trends
  \Statex

  \State \textit{Optimization Stage}
  \For{each rebalancing window $k$}
    \State Formulate mean–variance objective: 
          $\min_x \; q\,x^\top\Sigma x - \mu^\top x$
    \State Encode budget \& cardinality constraints: 
          $H = H_{\text{obj}} + \lambda H_c$
    \State Convert $H$ to QUBO using \texttt{QuadraticProgramToQubo}
    \State Solve using \texttt{SimulatedAnnealingSampler} 
          (\texttt{num\_reads = 5000})
    \State Decode lowest-energy bitstring and normalize to obtain $w^{(k)}$
  \EndFor
  \Statex

  \State \textit{Rebalancing and Evaluation}
  \For{each quarter $k \rightarrow k{+}1$}
    \State Rank current holdings by expected return $\mu$
    \State Sell the $K_{\text{sell}} = 4$ lowest-return assets to free capital
    \State Identify candidate replacements from the same sectors
    \State Re-solve the QUBO on this subset to select $K_{\text{sell}}$ new assets
    \State Allocate integer shares 
    \State Update holdings and compute portfolio value, risk, return, and Sharpe ratio
  \EndFor
\end{algorithmic}
\end{algorithm}

\begin{figure}[h!]
    \centering
    \includegraphics[width=0.5\textwidth]{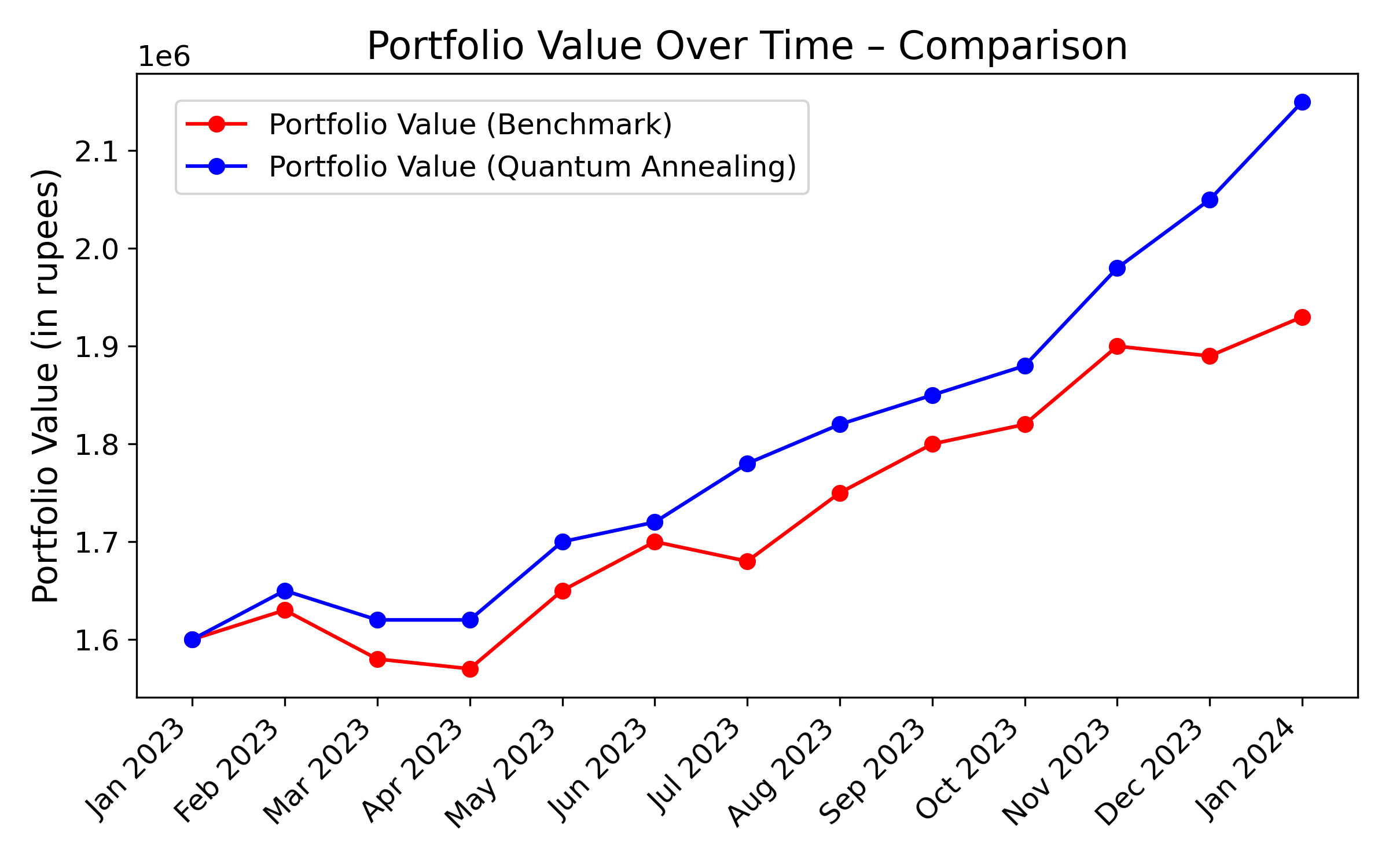}
    \caption{Comparison of both the portfolios }
    \label{fig:portfolio_comparison}
\end{figure}

In Fig. \ref{fig:portfolio_comparison}, the red line represents the portfolio value of the benchmark data for our analysis. Conversely, the blue line depicts the portfolio value achieved by our algorithm. Over the 13-month period, both the individual investor and the algorithm started with the same portfolio value of 1.6 million Indian Rupees. By the end of this period, the investor’s portfolio had grown to approximately 1.95 million, reflecting an absolute gain of 0.35 million rupees, or about 21.9\% growth. In comparison, the algorithm’s portfolio reached 2.2 million rupees, marking a larger increase of 0.6 million, or 37.5\%. This means the algorithm not only delivered a higher return overall but also outperformed the investor by around 0.25 million rupees in absolute terms, which translates to a performance gap of roughly 15.6 percentage points.

Following the computation and comparison of portfolio values, we employ the rebalancing strategies described in Section \ref{subsec:rebalancing}. In this study, the portfolios were rebalanced quarterly, resulting in a total of four rebalancing actions over the 13-month portfolio period.

\begin{figure}[h!]
    \centering
    \includegraphics[width=0.5\textwidth]{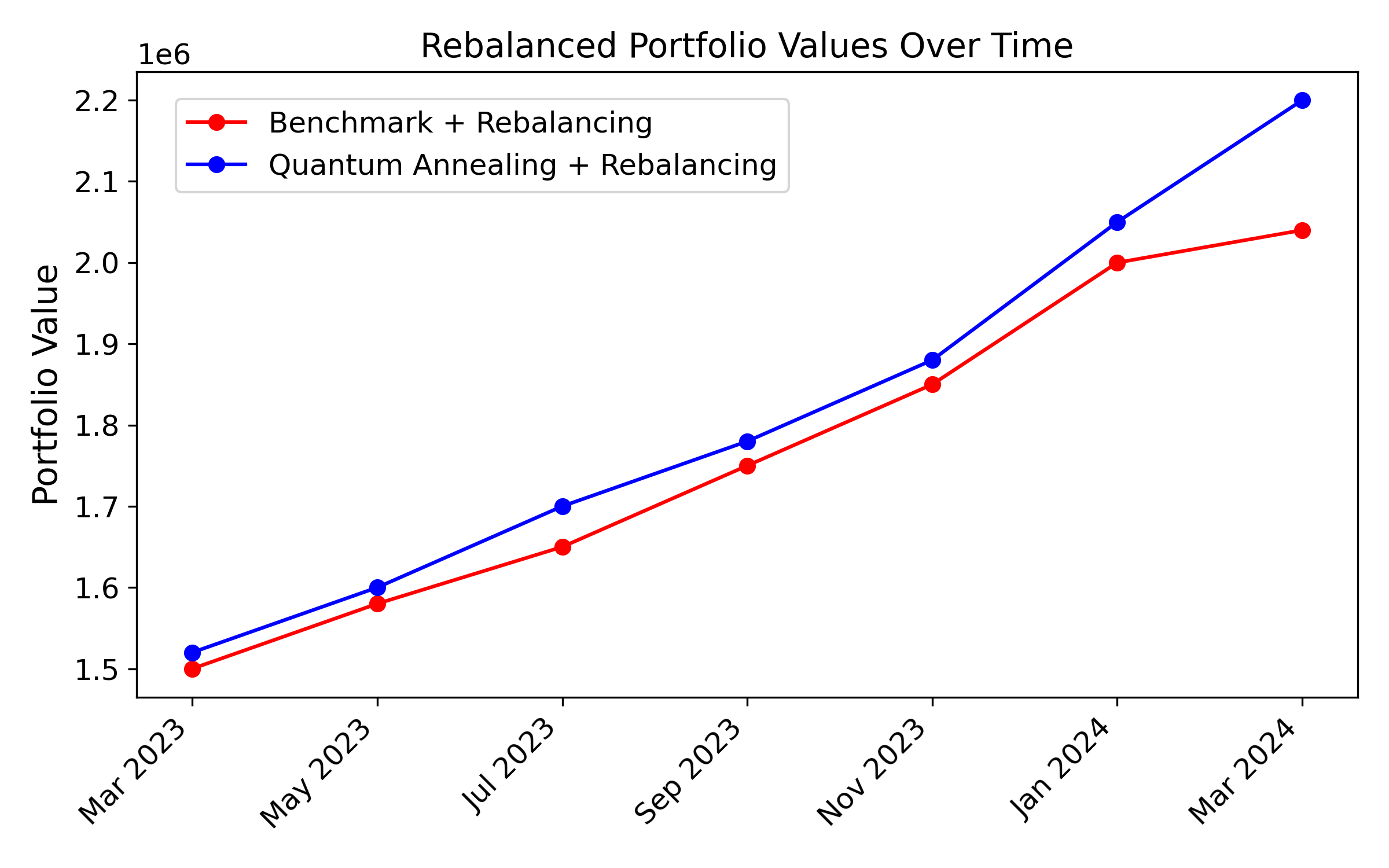}
    \caption{Comparison of both the portfolios after rebalancing as explained in \ref{subsec:rebalancing}.}
    \label{fig:rebalancing_comparison}
\end{figure}

Fig. \ref{fig:rebalancing_comparison} illustrates the rebalancing outcomes, with the red line representing the user's portfolio and the blue line depicting the algorithm's portfolio. After conducting four rebalancing sessions, the difference between the two portfolios becomes quite marginal. However, when contrasting these rebalanced portfolios with the original portfolios shown in Fig. \ref{fig:portfolio_comparison}, there is an observable improvement in their values. Specifically, the user's rebalanced portfolio surpasses 2 million Indian Rupees, whereas the original value was approximately 1.95 million Indian Rupees. This highlights the effectiveness of rebalancing in enhancing portfolio performance.

\begin{figure}[h!]
    \centering
    \includegraphics[width=0.5\textwidth]{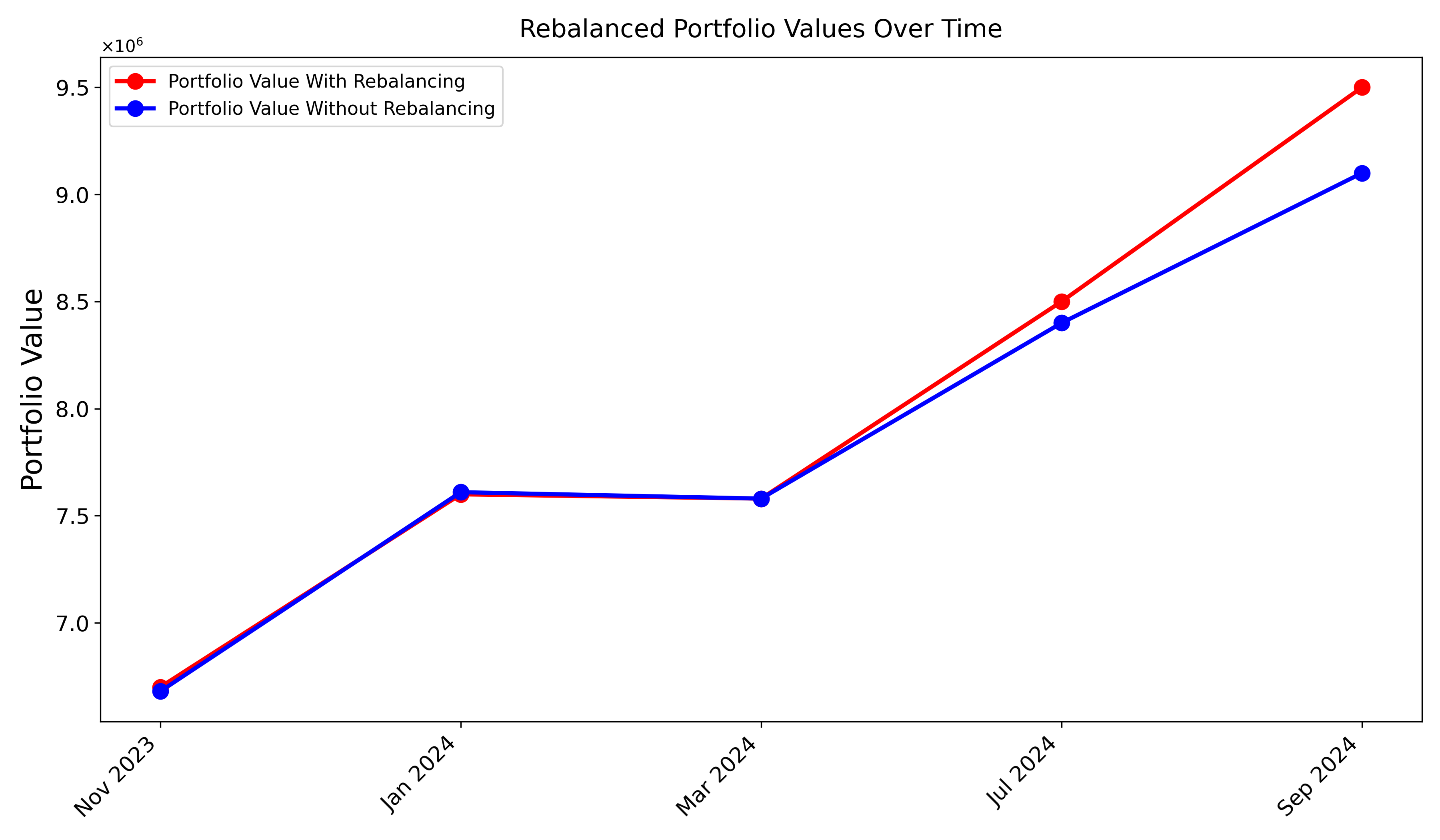}
    \caption{Comparison of a client portfolio before and after rebalancing}
    \label{fig:rebalancing_iciciC_nifty50}
\end{figure}

We did not restrain from testing the above methods with one single portfolio. Below are more results on different portfolios with various benchmarks used in the Indian market. In fig. \ref{fig:rebalancing_iciciC_nifty50} we apply the rebalancing technique to a portfolio by a real user which consists of 70 assets, where the user bought 100 shares of each asset. He started his portfolio on the 1st of September 2023 and stopped all his transactions on 25th September 2024. The benchmark used in this portfolio is NIFTY 50 which contains 50 of the largest Indian companies listed on National Stock Exchange(NSE).  

\begin{figure}[h!]
    \centering
    \includegraphics[width=0.5\textwidth]{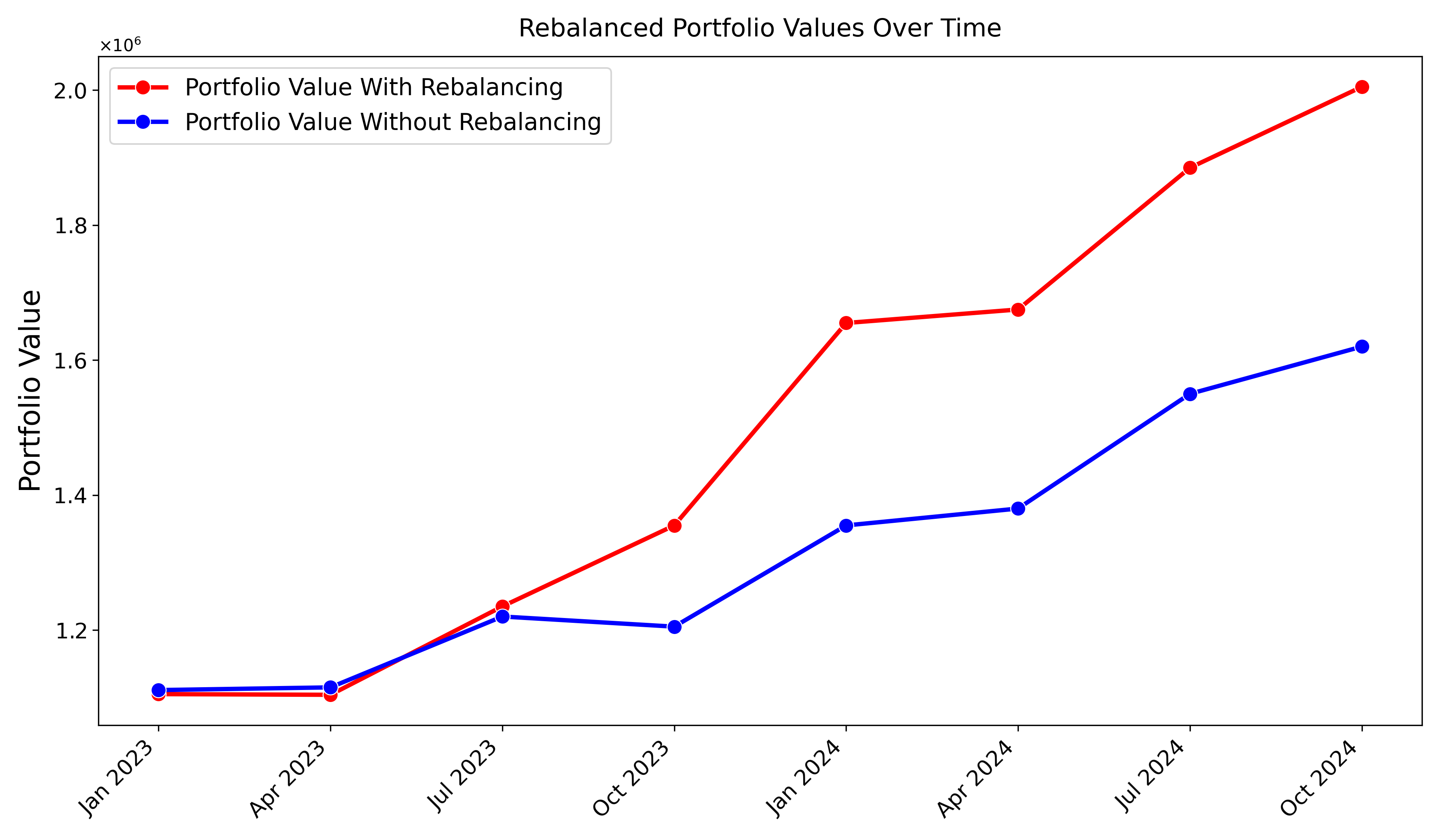}
    \caption{Comparison of a test portfolio before and after rebalancing with NIFTY 50 benchmark }
    \label{fig:rebalancing_test_nifty50}
\end{figure}

\begin{figure}[h!]
    \centering
    \includegraphics[width=0.5\textwidth]{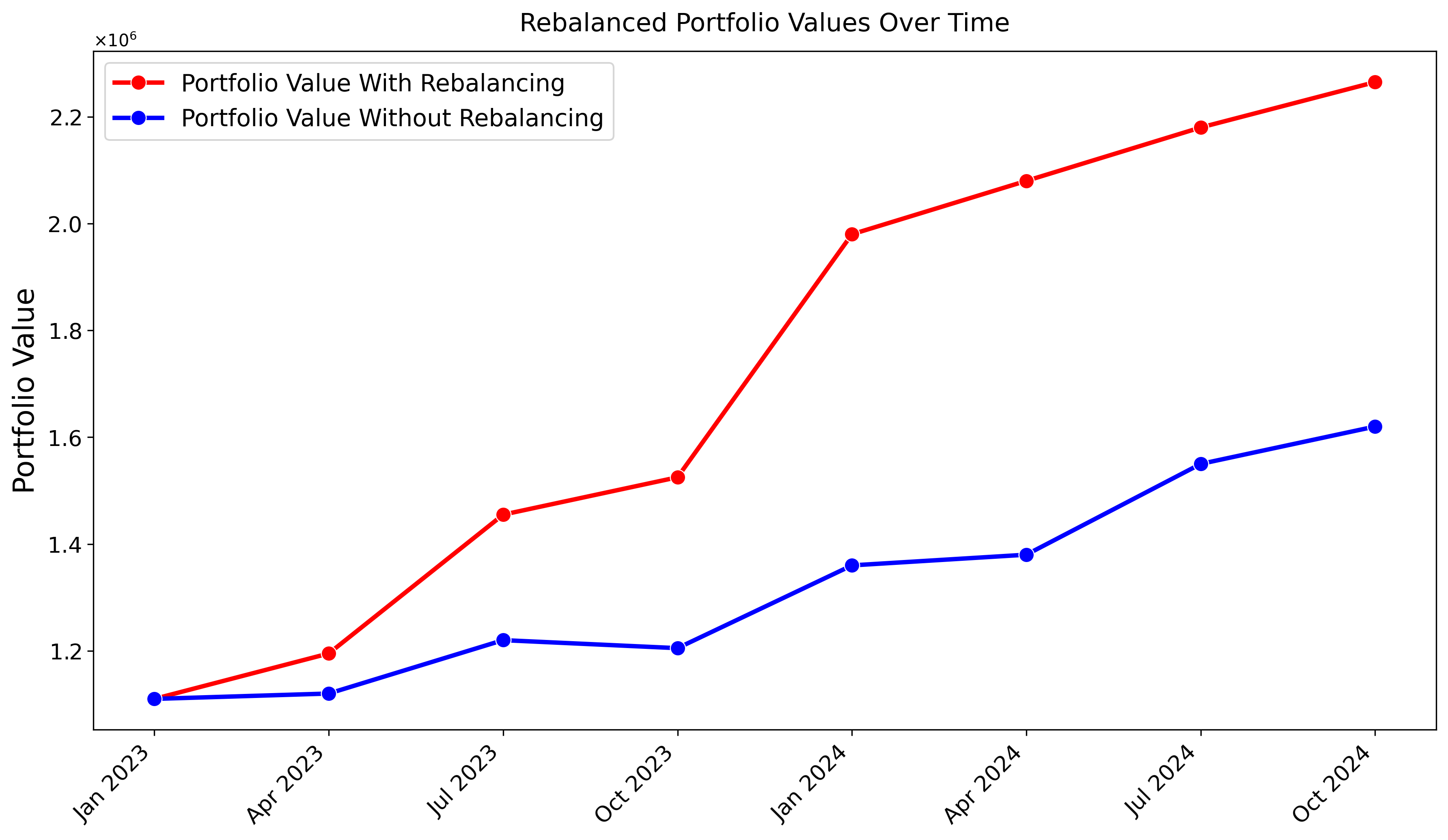}
    \caption{Comparison of a test portfolio before and after rebalancing with NIFTY 100 benchmark }
    \label{fig:rebalancing_test_nifty100}
\end{figure}

\begin{figure}[h!]
    \centering
    \includegraphics[width=0.5\textwidth]{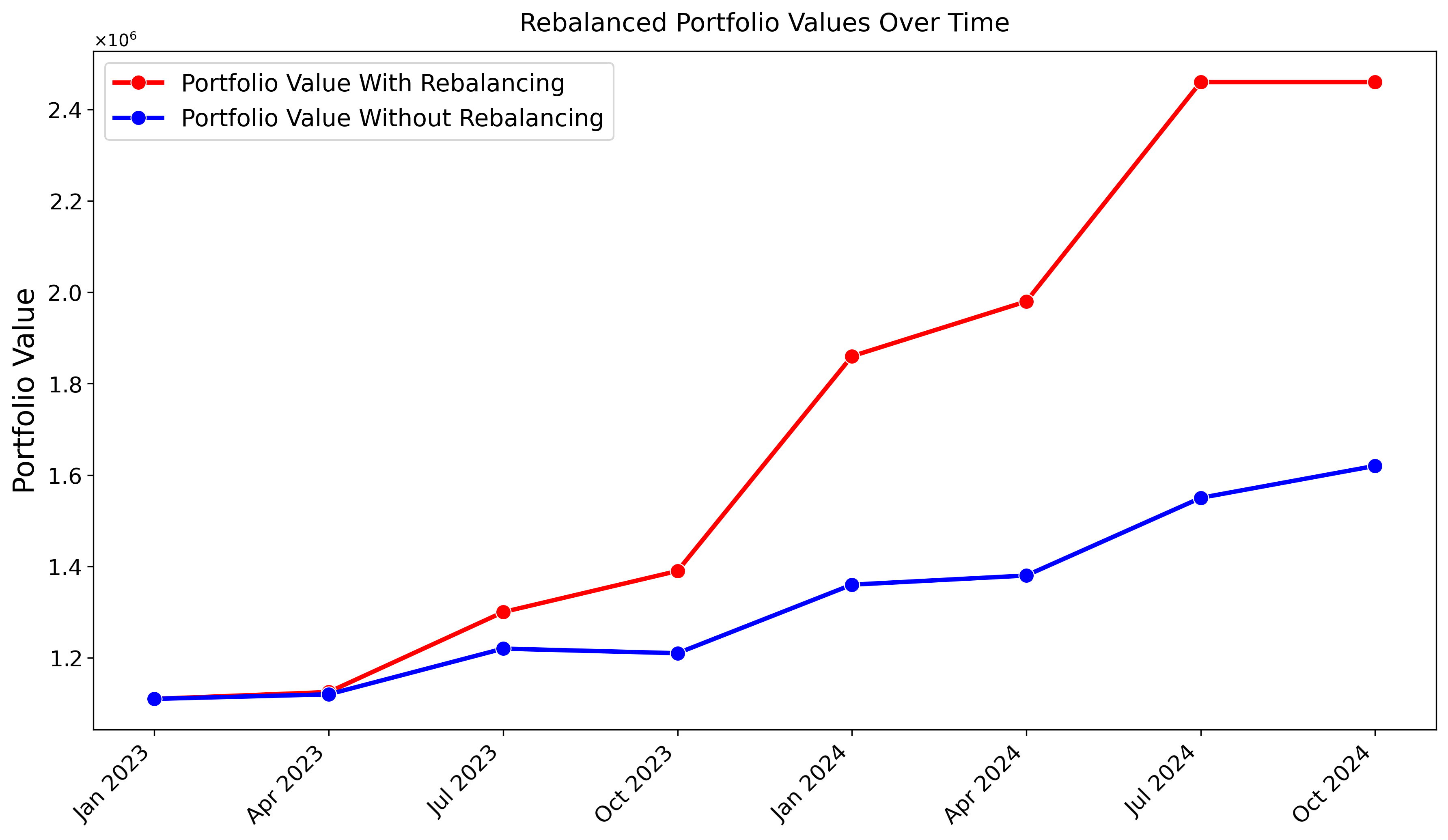}
    \caption{Comparison of a test portfolio before and after rebalancing with NIFTY 500 benchmark }
    \label{fig:rebalancing_test_nifty500}
\end{figure}

\begin{figure}[h!]
    \centering
    \includegraphics[width=0.5\textwidth]{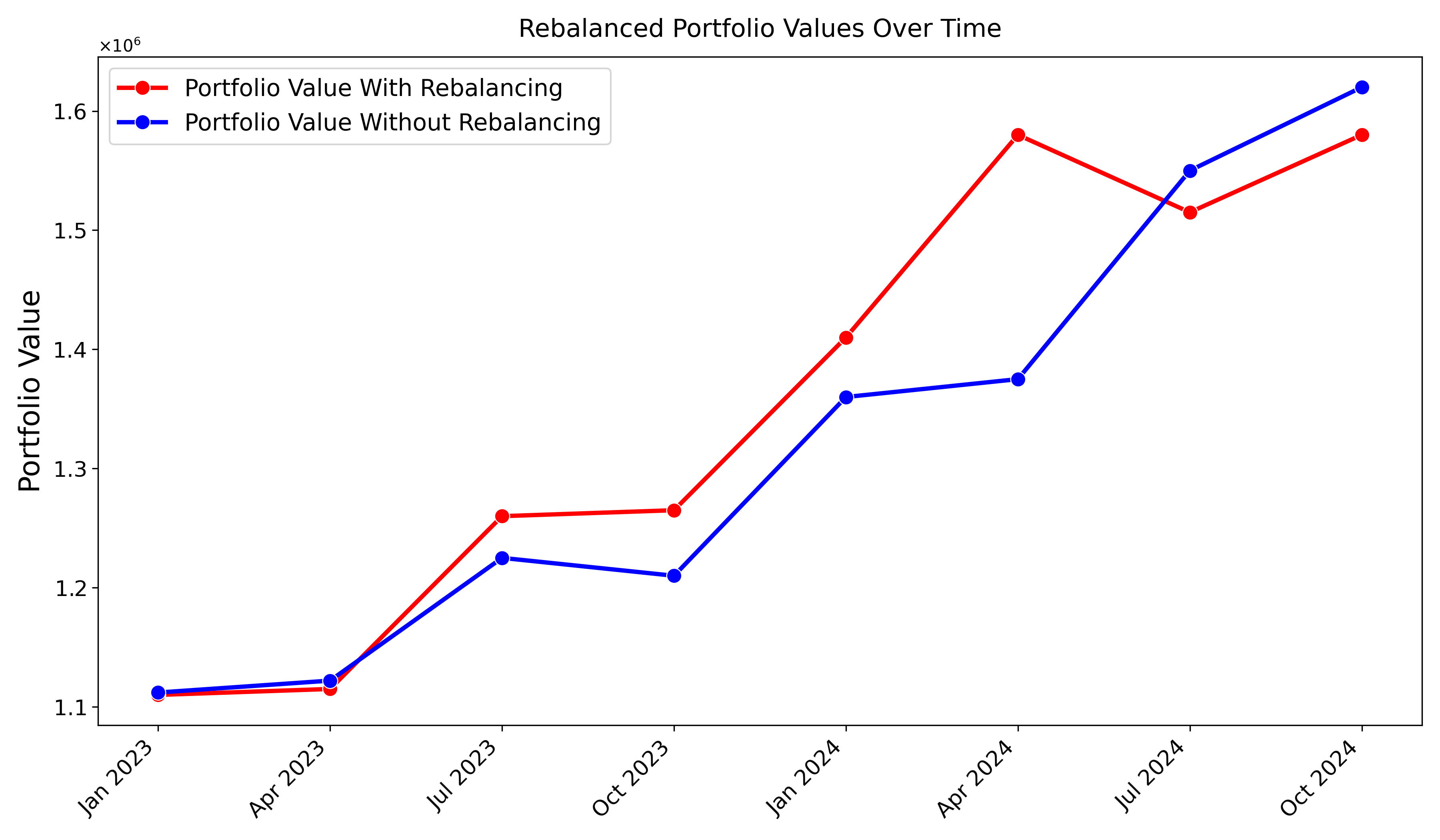}
    \caption{Comparison of a test portfolio before and after rebalancing with BSE 100 benchmark }
    \label{fig:rebalancing_test_bse100}
\end{figure}

\begin{figure}[h!]
    \centering
    \includegraphics[width=0.5\textwidth]{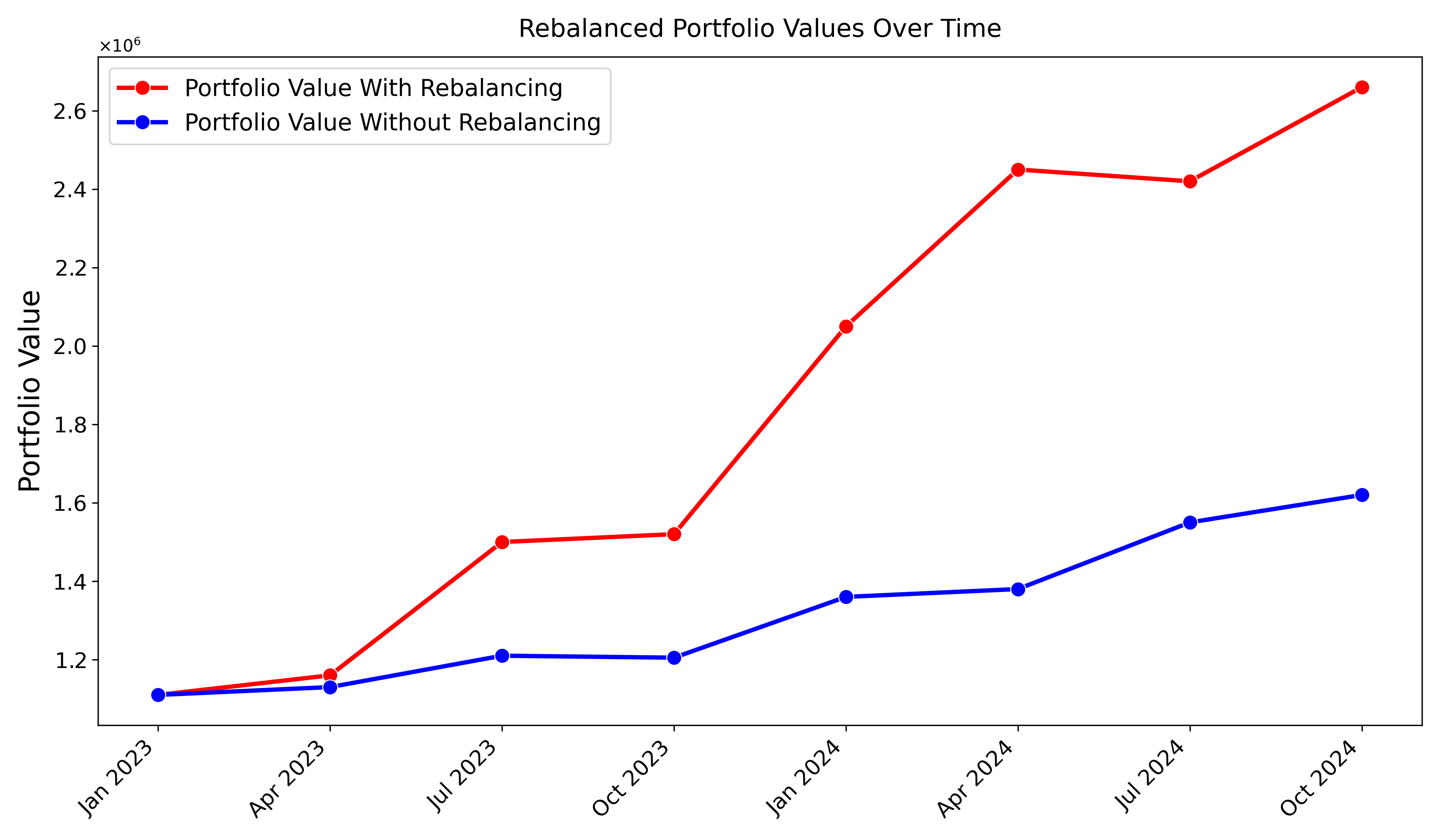}
    \caption{Comparison of a test portfolio before and after rebalancing with BSE 500 benchmark }
    \label{fig:rebalancing_test_bse500}
\end{figure}

We also created a test portfolio by ourselves which is not real rather fabricated. This synthetic portfolio contains of 11 assets from different sectors, Information technology, Health Care, Banking to name a few. We also assumed that there are 100 shares of each of these assets in this Test Portfolio for simplicity. We used various benchmarks to see the efficiency of our rebalancing technique. Fig. \ref{fig:rebalancing_test_nifty50}, \ref{fig:rebalancing_test_nifty100}, \ref{fig:rebalancing_test_nifty500}, shows the the impact of our rebalancing methodology with NIFTY 50, 100, 500 benchmarks respectively. Where as Fig. \ref{fig:rebalancing_test_bse100}, \ref{fig:rebalancing_test_bse500}, shows the performance with BSE 100, 500 benchmarks respectively.

\section{Conclusions}\label{sec:conc}

% \textcolor{blue}{
This work presents an end-to-end framework for portfolio optimization 
that integrates classical mean–variance modeling with quantum annealing-based 
rebalancing. The experiments demonstrate that, even under current hardware constraints, 
quantum annealers can successfully handle constrained portfolio formulations 
and yield allocations comparable to classical baselines.
% } 

% \textcolor{blue}{
However, the scope of the present study remains limited to a modest asset universe 
and should be interpreted as an initial feasibility test rather than a demonstration 
of large-scale deployment. 
The results are influenced by inherent limitations of current annealing technology, 
including restricted problem size due to embedding overhead, limited parameter precision, 
and the need for hybrid solver orchestration that introduces classical latency. 
Furthermore, the stochastic nature of the annealing process may lead to solution variability, 
necessitating multiple runs and statistical averaging.

% \textcolor{blue}{
These constraints imply that the current results reflect the near-term capabilities 
of quantum annealers rather than their ultimate performance potential. Given the strength of classical MIQP and metaheuristics (GA/SA), our results should be interpreted as evidence of feasibility rather than superiority. Future research will aim to benchmark larger portfolios, explore advanced hybrid workflows, 
and leverage improvements in annealing hardware and problem embedding to approach 
industrial-scale portfolio optimization.

\section*{Future Work}
While this study demonstrates the feasibility of integrating quantum annealing into a practical portfolio construction pipeline, several important avenues remain open for future exploration.
The present experiments use a modest asset universe motivated by the constraints of our industry collaboration and the hybrid solver configurations available at the time of execution. Extending the analysis to broader such as the global equity indices, or multi-asset datasets will be essential for assessing scalability and stress-testing the proposed pipeline under realistic institutional workloads.
The current quarterly rebalancing procedure can be extended by explicitly modeling transaction costs, slippage, and liquidity constraints.  Incorporating these frictions will produce a more realistic assessment of turnover and its impact on realized performance. Given evolving access policies for commercial annealers, future work will explore alternative quantum and quantum-inspired backends, including simulated quantum annealers, Fujitsu DAU, Toshiba SBM, and emerging open-access annealing platforms. 
This will allow the pipeline to be tested under diverse hardware characteristics and will facilitate reproducibility.
Overall, this work provides an initial proof-of-concept integration of quantum annealing into a portfolio-optimization workflow.  Future efforts will focus on broadening empirical scope, strengthening baselines, improving reproducibility, and evaluating scalability across markets, datasets, and hardware platforms.

\section*{Acknowledgment}
The authors would like to thank D-Wave Systems for the access to their quantum annealer. We would also like to thank Prof. Greg Byrd for the meaningful discussions and comments. This work is supported by the NSF, Office of Strategic Initiatives, under Grant No. OSI-2328774 (SM,KI)

% \textcolor{blue}{
\section*{Data and code Availability} All code and processed datasets required to reproduce the results will be provided by the authors upon reasonable request. 

\bibliographystyle{IEEEtran}  
\bibliography{IEEEabrv,References}  
 
\end{document}